\documentclass[conference]{IEEEtran}
\IEEEoverridecommandlockouts
\usepackage{cite}
\usepackage{amsmath,amssymb,amsfonts}
\usepackage{algorithmic}
\usepackage{graphicx}
\usepackage{textcomp}
\usepackage{xcolor}
\usepackage{amsmath,epsfig}
\usepackage{cite}
\usepackage{amssymb}
\usepackage{amsmath}
\usepackage{graphicx}
\usepackage{subfigure}
\usepackage{epstopdf}
\usepackage{setspace}
\usepackage{booktabs}
\usepackage{multirow}
\usepackage{gensymb}
\usepackage{makecell}
\usepackage{amssymb}
\usepackage{cite}
\usepackage{amssymb}
\usepackage{amsmath}
\usepackage{epsfig}
\usepackage{subfigure}
\usepackage{epstopdf}
\usepackage{setspace}
\usepackage{booktabs}
\usepackage{multirow}
\usepackage{makecell}
\usepackage{gensymb}

\def\BibTeX{{\rm B\kern-.05em{\sc i\kern-.025em b}\kern-.08em
    T\kern-.1667em\lower.7ex\hbox{E}\kern-.125emX}}
\begin{document}

\title{A deep deformable residual learning network for SAR images segmentation\\
\thanks{This work was supported by the National Natural Science Foundation of China under Grants 61901091 and 61671117}
}

\author{\IEEEauthorblockN{Chenwei Wang, Jifang Pei, Yulin Huang, Jianyu Yang}
\IEEEauthorblockA{\textit{School of Information and Communication Engineering} \\
\textit{University of Electronic Science and Technology of China}\\
Chengdu, China \\
Email: dbw181101@163.com}
}

\maketitle

\begin{abstract}
Reliable automatic target segmentation in Synthetic
Aperture Radar (SAR) imagery has played an important role in the
SAR fields. Different from the traditional methods,
Spectral Residual (SR) and CFAR detector, with the recent adavance in machine learning theory,
there has emerged a novel method for SAR target segmentation, based on the deep learning
networks. In this paper, we proposed a deep deformable residual learning network
for target segmentation that attempts to preserve the precise contour of the target.
For this, the deformable convolutional layers and residual learning block are applied, which could extract and preserve the geometric information of the targets as much as possible.
Based on the Moving and Stationary Target Acquisition and Recognition (MSTAR) data set, experimental results have shown the superiority of the proposed network for the precise targets segmentation.
\end{abstract}


\section{Introduction}
Synthetic aperture radar (SAR) is a type of imaging radar which can obtain high resolution image under very low visibility conditions.
As a coherent imaging sensor, with the ability to operate at night and in adverse weather
conditions such as thin clouds and haze, SAR has gradually become a significant source of remote
sensing data in the fields of geographic mapping, resource surveying, and military reconnaissance \cite{wang2022recognition,wang2023entropy,wang2023sar,wang2022global,wang2022semi,wang2020deep,wang2022sar,wang2021multiview,wang2019parking,wang2020multi}.
The geometric features of SAR images could also provide distinctive information for the image interpretation.
Therefore, SAR image segmentation is an essential
processing step and has become one of
the most crucial and challenging issues in SAR fields.

The segmentation of SAR images plays a very important role in the interpretation and understanding of SAR images.
It has received an increasing amount of attention.
However, different from optical images, caused by the coherent nature of the scattering phenomena \cite{rosen2000synthetic}, SAR images are inherently affected by multiplicative noise.
The noise affects the quality of SAR images and the performance of SAR image segmentation.

A number of segmentation methods of SAR images have been proposed since decades.
These methods include graph partitioning techniques \cite{shi2000normalized}, clustering algorithms \cite{zhang2008spectral}, model based
methods \cite{descombes1996coastline}, and morphologic strategies \cite{lemarechal1998sar}, many improved fuzzy
C-means \cite{pham1999adaptive}.
However, with not considering any information about spatial context, these methods are very sensitive to noise and other imaging artifacts.
Furthermore, all the above methods are based on the hand-engineered features,which might be robust in the practical applications.

Deep learning methods, which learn the hierarchical representative features, have recently
become focus SAR image automatic interpretation.
In recent years, deep-learning-based methods for SAR image segmentation have proved promising and state-of-art,
such as convolutional neural networks for SAR image segmentation \cite{liu2016sar}, \cite{duan2017sar}.
However, the segmentation results of these methods might loss the meticulous contour structure,
which means that the sharp feature and details of the images will fade off through the pipe of the processes.

In this paper, A deep deformable residual learning network is proposed for SAR images segmentation.
The network firstly extracts the image features by employing the deformable convolutional layers.
These layers could change the shape of the kernels based on the targets' shape in the images and preserve the information of contour of targets.
Then, we employ the residual learning blocks to preserve the extracted crucial features with the depth going higher and accomplish the target segmentation from SAR images.
Therefore, the network has the capability of extract the optimal features and the sharp contour from the SAR images and achieved the superior performance of segmentation in SAR images.

The rest of the paper is organized as follows. In Section II, the network and crucial configuration will be present.
In Section III, experimental results are provided. In Section IV, the final conclusion is summarized.

\section{the proposed method}
In this section, the architecture and two crucial configurations of the proposed network will be described in detail. First, we elucidate the specific structure of the architecture. Then, the deformable convolutional layer and the residual learning block will be present respectively.

\subsection{architecture}

As mentioned above, the crucial features and geometric structure
of the targets will gradually fade off partly through the pipe of the process.
Therefore, we propose a deep deformable residual learning network to efficiently extract
the contour information of the targets with preserving the information in the process
to achieve superior performance of SAR images segmentation.

The proposed deformable residual learning network is mainly
consist of two parts: the feature extractor and reconstructor, as shown in Fig.1.

The feature extractor is consist of three residual learning block \cite{he2016deep} to extract and
preserve the geometric and scattering information. The residual learning block is consist of
one deformable convolutional layers and two normal convolutional layers. The one deformable convolutional layers could change the kernel shape to reserve the geometric information in the image and extract the
optimal feature. The two normal convolutional layers and the skip connections could help the reservation of the information extracted by the deformable convolutional layers and reduce the gradient vanishing.
The reconstructor is constructed to adopt the fusion of the extracted features and preserve the information which represent the overall contour and local details of the images. The structure of the reconstructor is simply another residual learning block with three normal convolutional layers.

Through the above structures, the proposed deep deformable residual learning network
can extract and reserve optimal features layer by layer from SAR images
and employ these extracted features adaptively and optimally
to achieve the superior performance of SAR image directed segmentation.

Loss functions form an important and integral part of learning
process, especially in CNN-based image reconstruction tasks. In this paper, we adopted the Euclidean distance as the loss as follow.

Given an image pair $\left\{ {X,Y} \right\}$, where $Y$ is the SAR
image and $X$ is the corresponding ground truth of the segmentation, the Euclidean loss is defined as
\begin{align}
\label{rangereso1}
Los{s_e} = \frac{1}{{WH}}\sum\limits_{w = 1}^W {\sum\limits_{h = 1}^H {\left\| {net\left( {{Y_{w,h}}} \right) - {X_{w,h}}} \right\|} _2^2}
\end{align}
\begin{figure}[!htb]
\centering
\centering\includegraphics[width=0.5\textwidth]{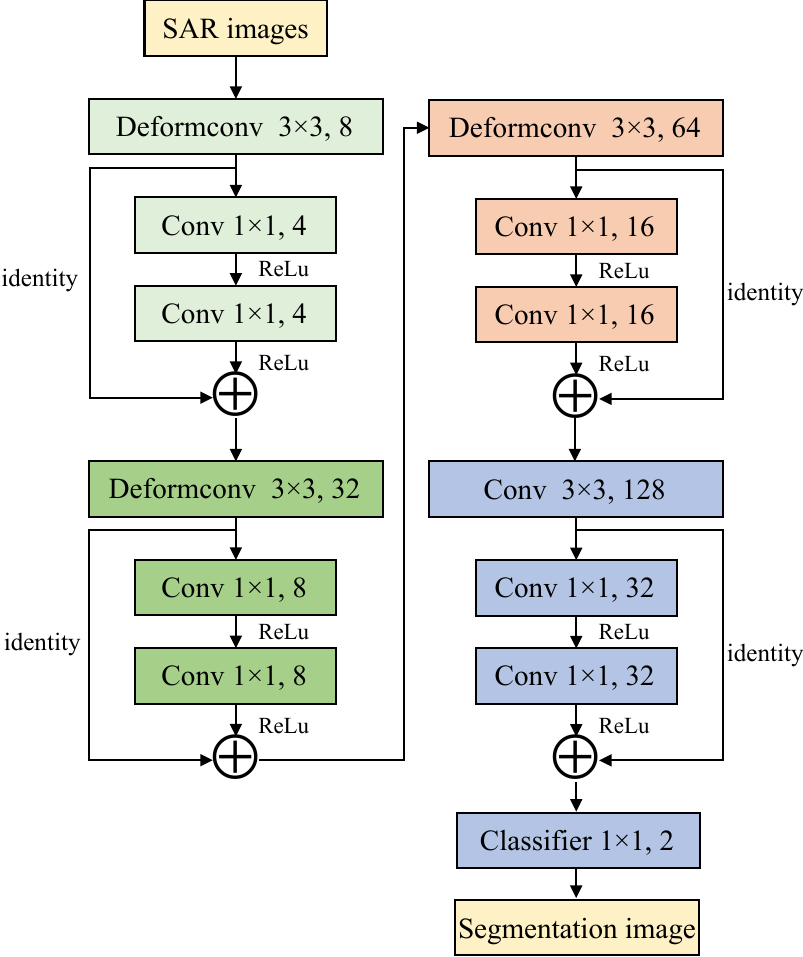}
\caption{The proposed deep deformable residual learning network for SAR images segmentation.}\label{flowchart1}
\end{figure}
\subsection{deformable convolutional layer}
The deformable convolutional layer proposed by \cite{dai2017deformable} could enhance CNN's capability of modeling geometric transformations . This new module adds 2D offsets to the regular grid sampling locations in the standard
convolution. It enables free form deformation of the sampling grid.
The offsets are learned from the preceding feature maps, via additional
convolutional layers. Thus, the deformation is conditioned
on the input features in a local, dense, and adaptive manner.
The stages of the 2D deformable convolutional layers could be described as follow.
First, sampling over the input feature map ${\bf{x}}$ using the offset;
\begin{align}
\label{rangereso1}
{\bf{x}}\left( {\bf{p}} \right) = \sum\limits_{\bf{q}} {G\left( {{\bf{q}},{\bf{p}}} \right) \bullet } {\bf{x}}\left( {\bf{q}} \right)
\end{align}
where ${\bf{p}}$ denotes an arbitrary (fractional) location, ${\bf{q}}$ enumerates all integral spatial
locations in the feature map ${\bf{x}}$, and $G\left(  \bullet  \right)$ is the bilinear interpolation kernel.

Then,summation of sampled values weighted by convolutional kernels ${\bf{w}}$.
\begin{align}
\label{rangereso1}
{\bf{y}}\left( {{{\bf{p}}_0}} \right) = \sum {\bf{w}} \left( {{{\bf{p}}_n}} \right) \cdot {\bf{x}}\left( {{{\bf{p}}_0} + {{\bf{p}}_n} + \Delta {\bf{p}}} \right)
\end{align}
where ${\bf{y}}$ denotes the output feature maps, ${{{\bf{p}}_0}}$ denotes the each location on ${\bf{y}}$.

\section{EXPERIMENTAL RESULTS}
In this section, we present the results of our proposed segmentation methods. The experiments are carried out on MSTAR dataset. First, the dataset will be introduced briefly, the segmentation notation is present and the network set-up will be described. Then, the experimental results will be present visually and qualitatively.

\subsection{Dataset and Network Set-up}
\begin{figure*}[!htb]
\centering
\centering\includegraphics[width=1\textwidth]{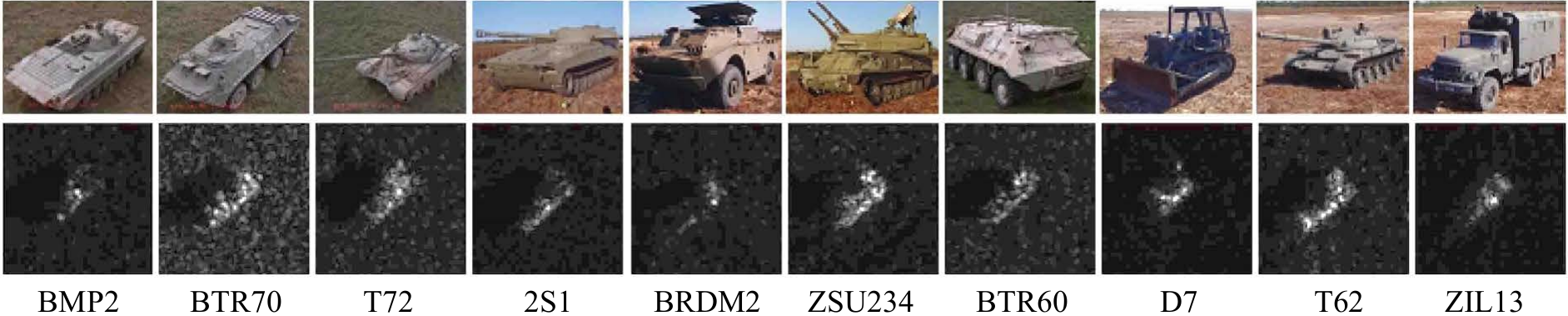}
\caption{SAR images and corresponding optical images of targets at similar aspect angels. From left to right: BMP2, BTR70, T72, 2S1, BRDM2, ZSU234,
BTR60, D7, T62, ZIL131.}\label{flowchart1}
\end{figure*}

\begin{table*}[!t]
\renewcommand{\arraystretch}{1.3}
\caption{Number of training and testing images for SOC before the data augmentation.}
\centering
\begin{tabular}{p{3.5cm}<{\centering}|p{2.6cm}<{\centering}|p{2.6cm}<{\centering}|p{2.6cm}<{\centering}|p{2.6cm}<{\centering}}
\hline \hline
  & \multicolumn{2}{c|}{Training} & \multicolumn{2}{c}{Testing}  \\
\hline
Class  & Depression & Number & Depression & Number  \\
\hline
BMP2-9563  & 17\degree & 233 & 15\degree & 196  \\
\hline
BTR70-c71 & 17\degree & 233 & 15\degree & 196  \\
\hline
T72-132 & 17\degree & 232 & 15\degree & 196  \\
\hline
BTR60-7532 & 17\degree & 256 & 15\degree & 195  \\
\hline
2S1-b01 & 17\degree & 299 & 15\degree & 274  \\
\hline
BRDM2-E71 & 17\degree & 298 & 15\degree & 274  \\
\hline
D7-92 & 17\degree & 299 & 15\degree & 274  \\
\hline
T62-A51 & 17\degree & 299 & 15\degree & 273  \\
\hline
ZIL131-E12& 17\degree & 299 & 15\degree & 274  \\
\hline
ZSU234-d08 & 17\degree & 299 & 15\degree & 274  \\
\hline \hline
\end{tabular}
\label{table1}
\end{table*}
The experiment dataset used to evaluate our proposed network is collected from the Moving
and Stationary Target Acquisition and Recognition (MSTAR)
program. The MSTAR data set  is released by the U.S
Defend Advanced Research Project Agency and the U.S Air
Force Research Laboratory. Ten different classes of ground targets
(tank: T62, T72, rocket launcher: 2S1, truck: ZIL131, armored
personnel carrier: BTR70, BTR60, BRDM2, BMP2, air defense
unit: ZSU23/4 and bulldozer: D7) are captured  using
the STARLOS sensor under X-band and contains a series of
0.3mx0.3m SAR images of ten different classes of ground targets.
The SAR images and responding optical images are presented
in Fig.2.
The training dataset is set as the images at ${17^ \circ }$, and the testing dataset is set as the images at ${15^ \circ }$. The numbers of the training and testing images are shown in Table I.
\begin{figure}[!htb]
\begin{center}
\subfigure[]{\label{1.1}\includegraphics[scale=0.1]{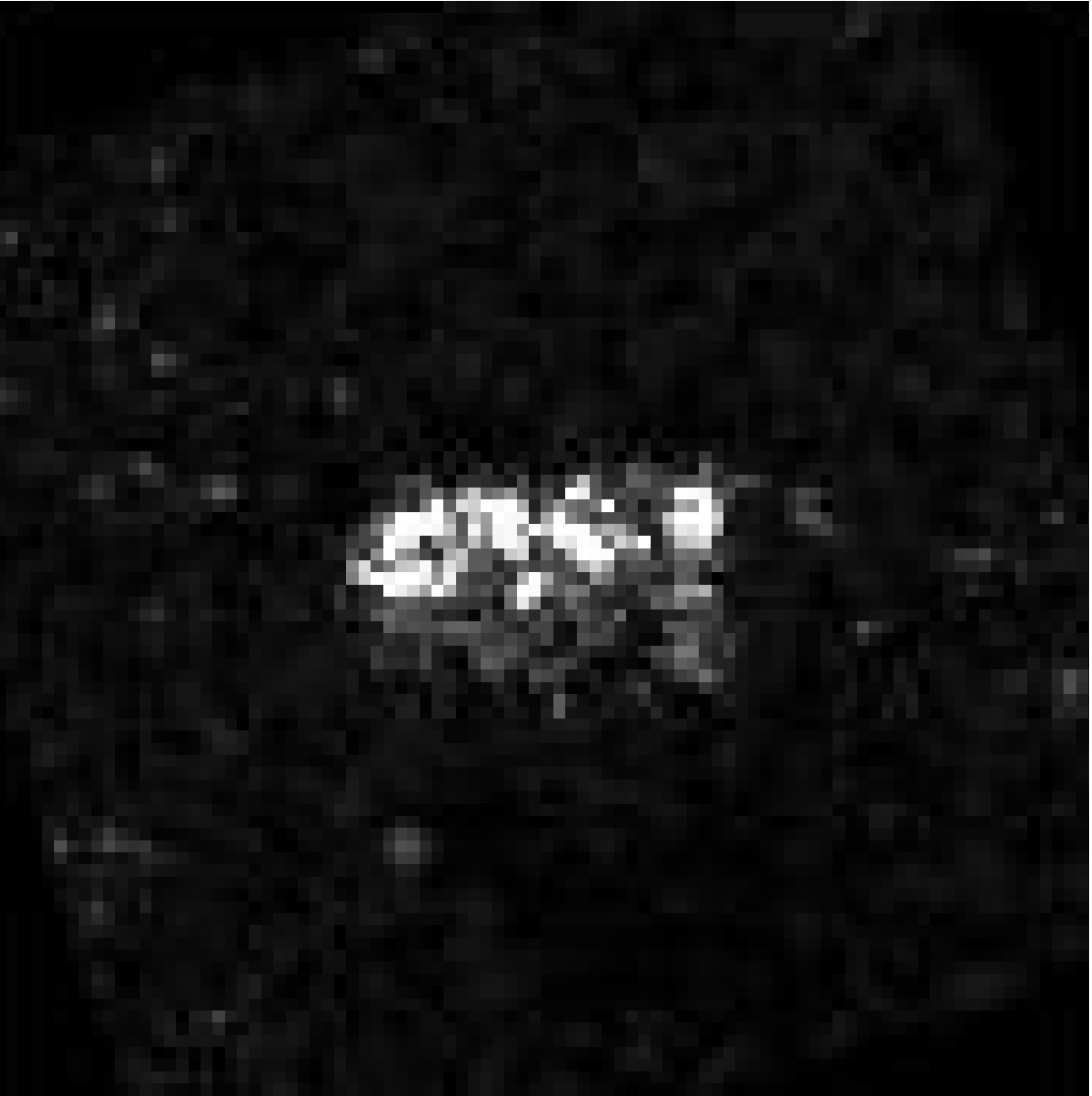}} \subfigure[]{\label{1.2}\includegraphics[scale=0.1]{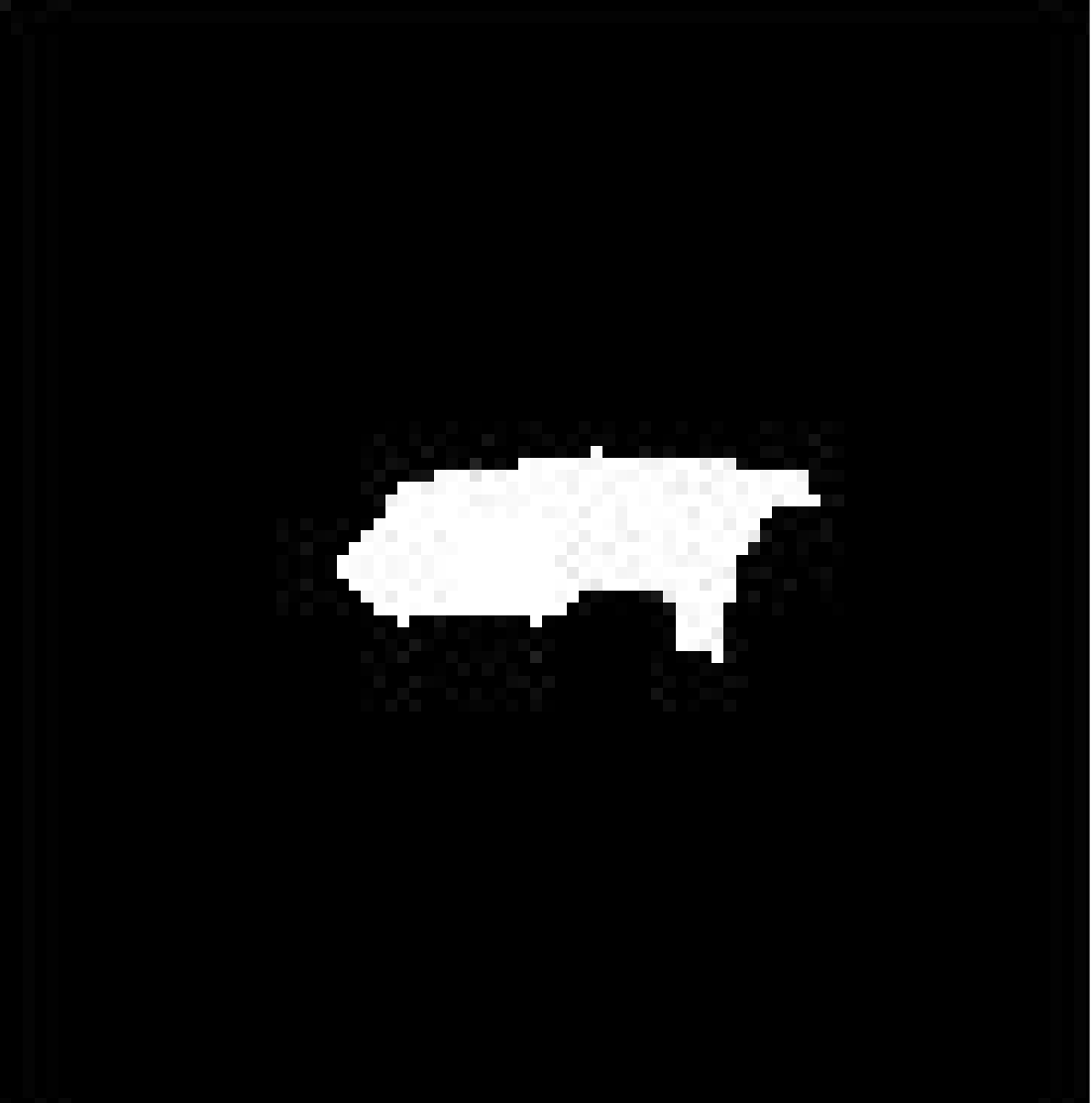}} \subfigure[]{\label{1.3}\includegraphics[scale=0.1]{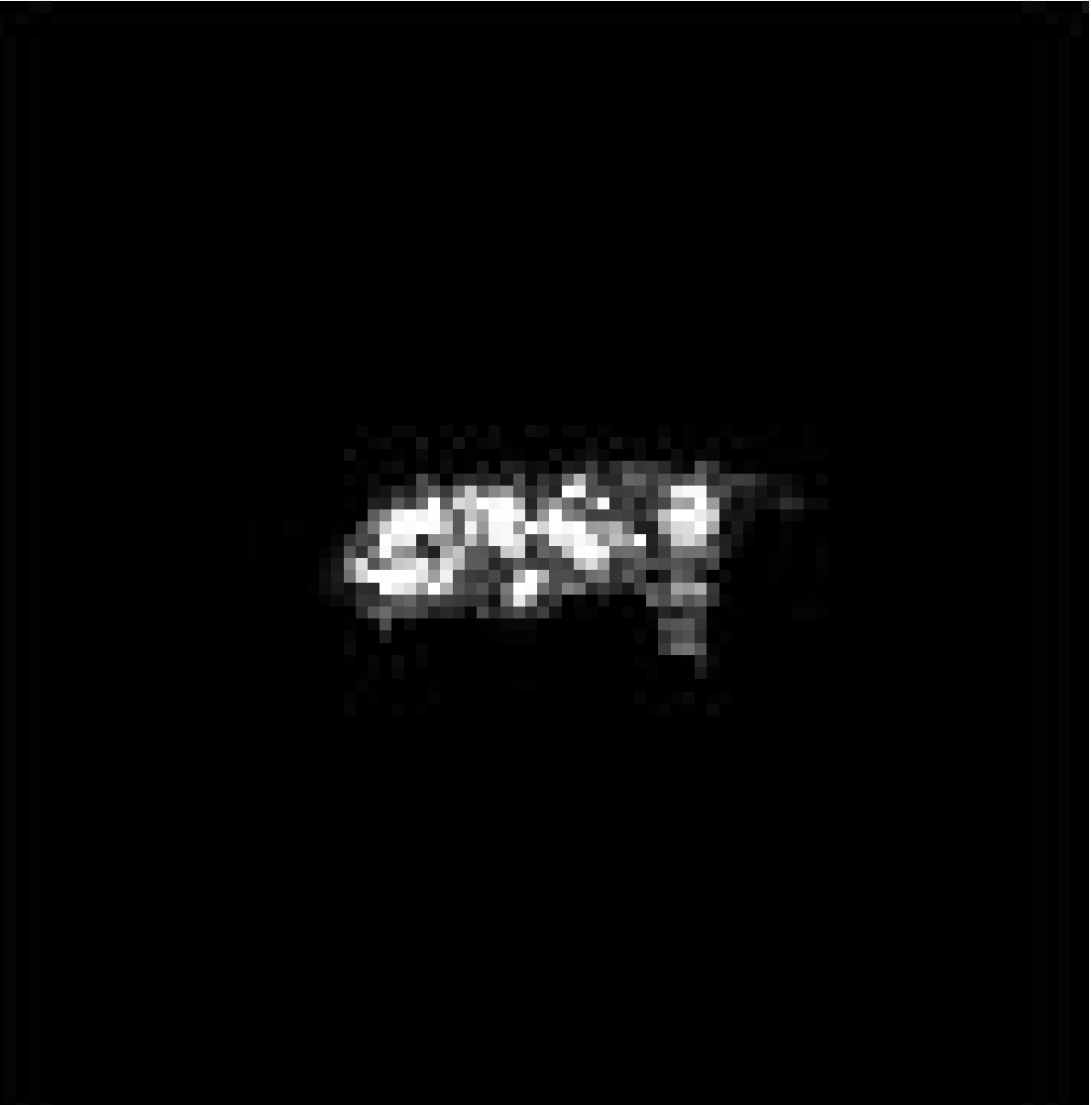}} \\ \subfigure[]{\label{2.1}\includegraphics[scale=0.1]{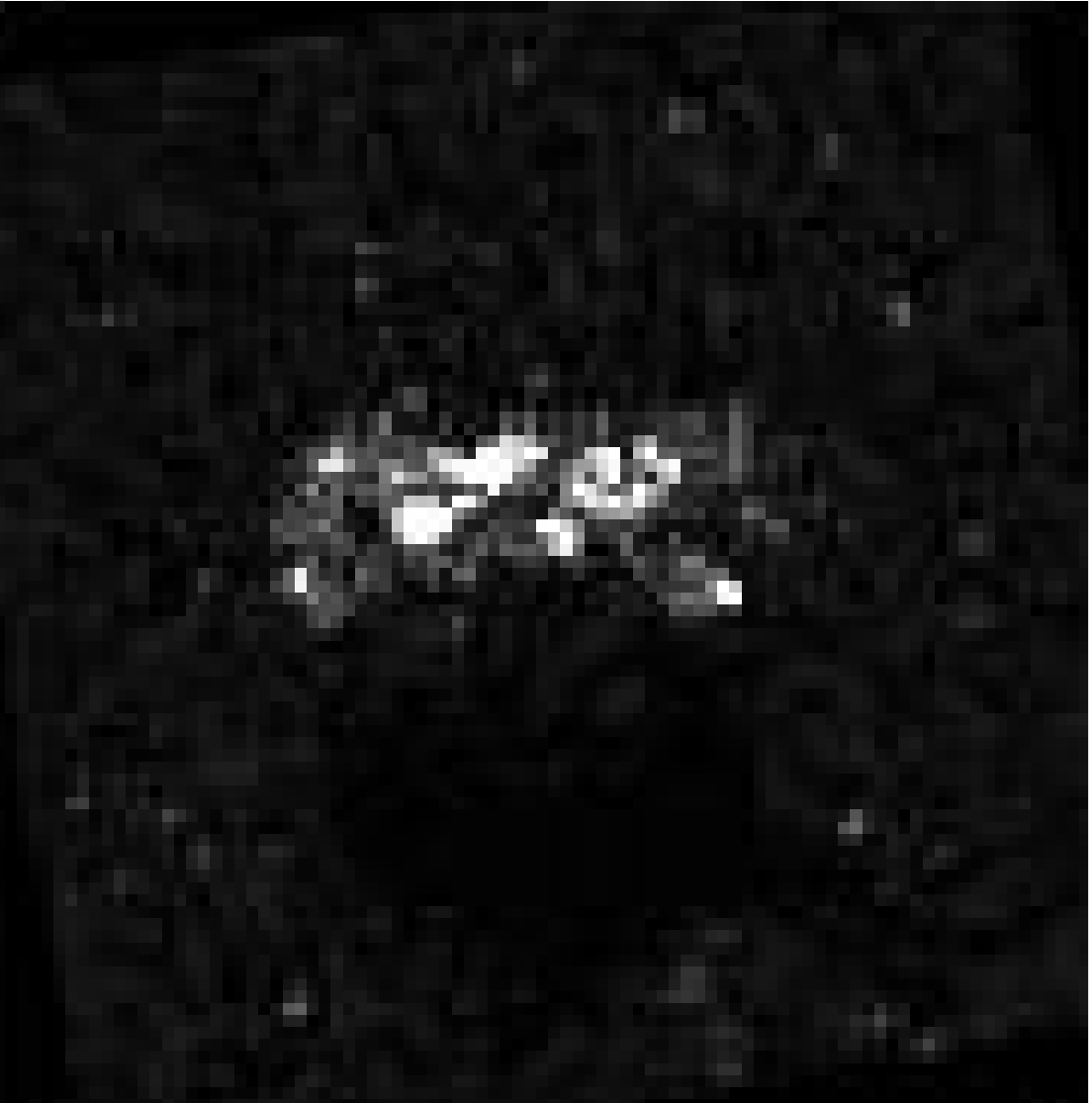}} \subfigure[]{\label{2.2}\includegraphics[scale=0.1]{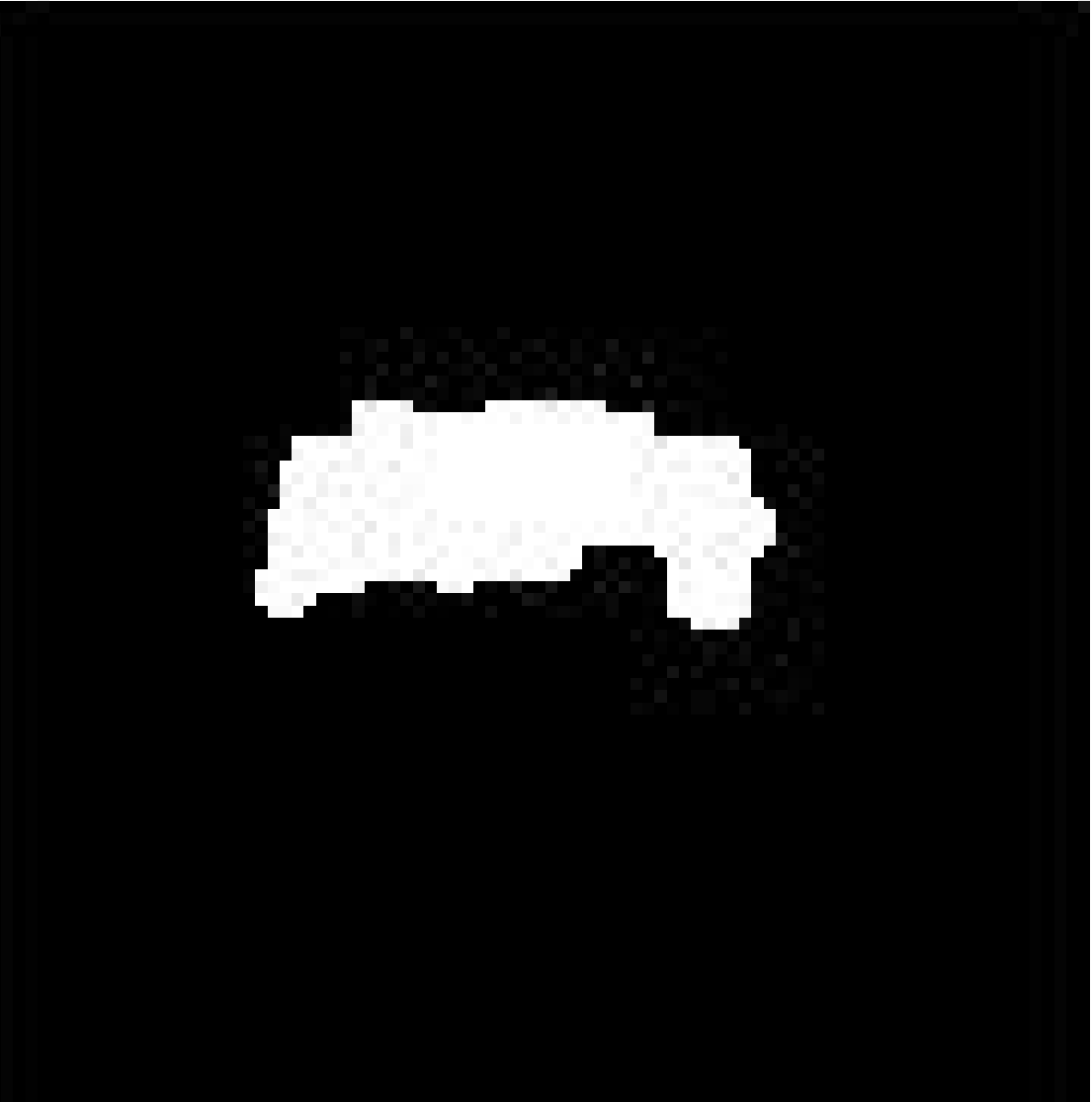}}  \subfigure[]{\label{2.3}\includegraphics[scale=0.1]{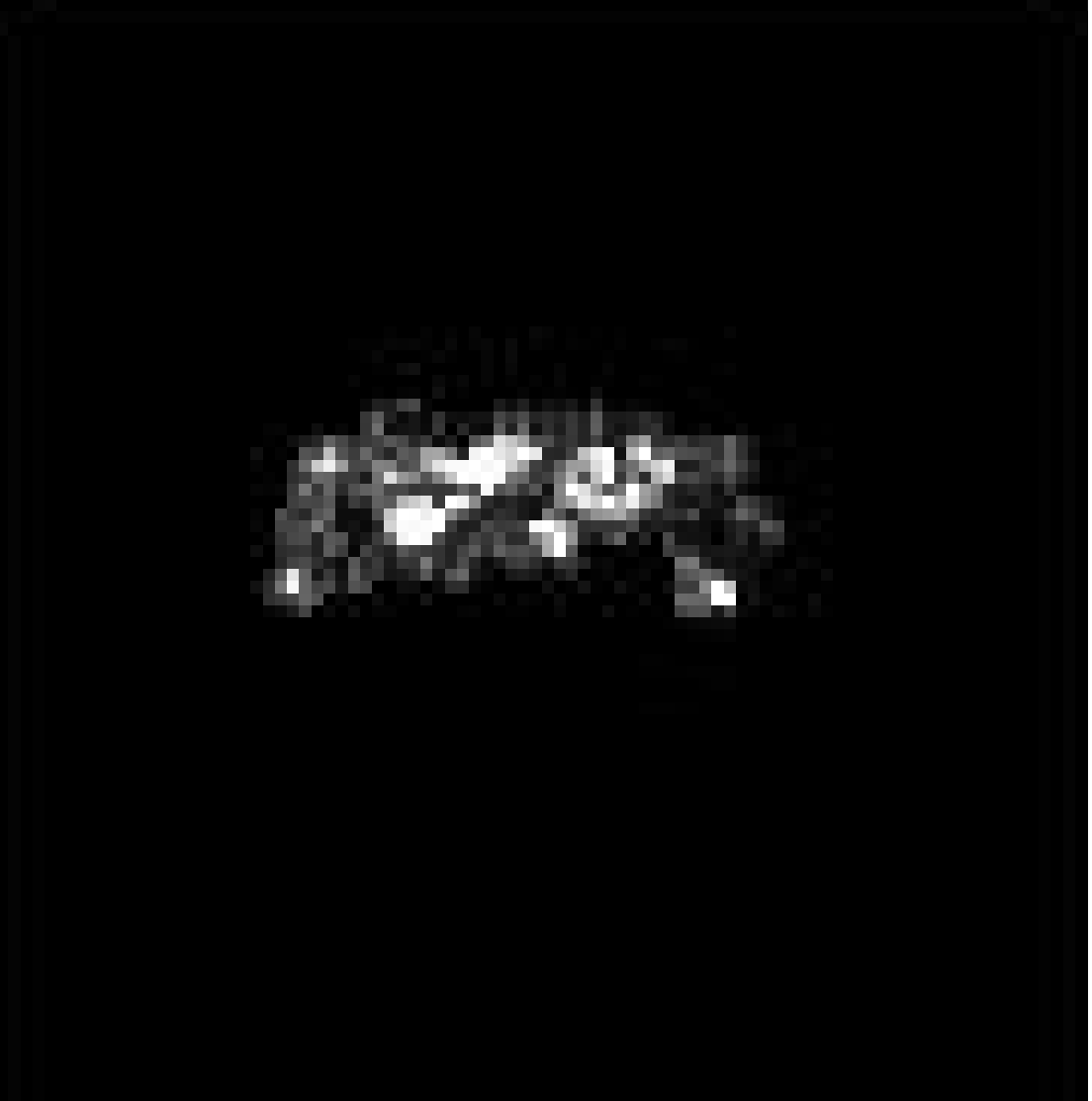}} \\
\end{center}
\caption{Some examples of the segmentation labels for different targets. First column: original SAR image, second column: segmentation ground truth, third column: masked original image.}
\label{multibeam}
\end{figure}

\begin{figure*}[!htb]
\begin{center}
\subfigure[]{\label{1.1}\includegraphics[scale=0.75]{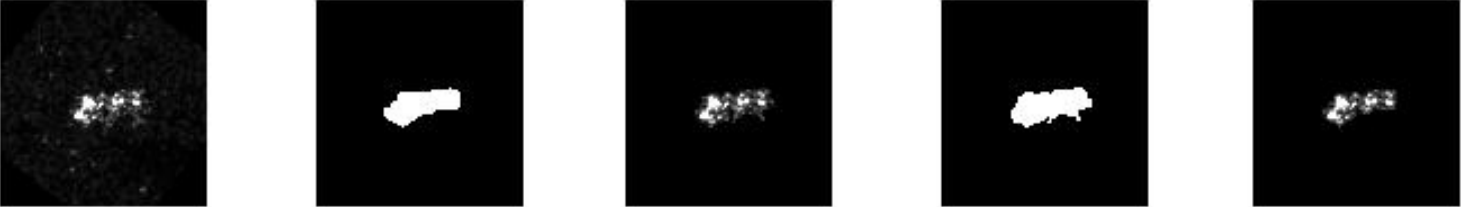}} \subfigure[]{\label{1.2}\includegraphics[scale=0.75]{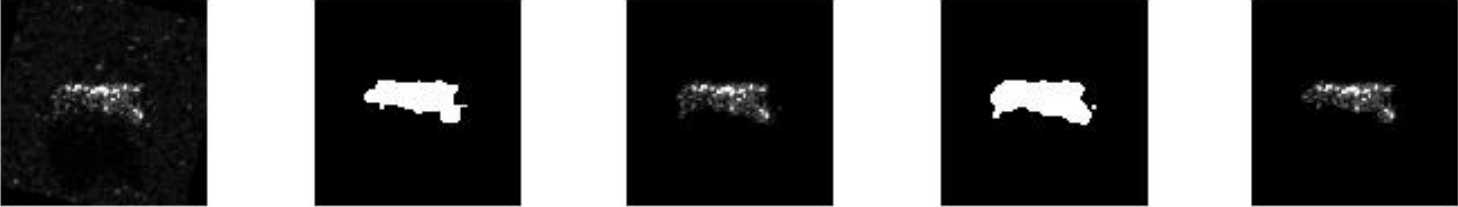}} \\
\end{center}
\caption{ Some segmentation results of the proposed network architecture for different targets.}
\label{multibeam}
\end{figure*}
Then, the training and
testing datasets of the segmentation are acquired by manual
annotation using the tool named OpenLabeling, as shown in Fig.3. The manual
annotation is based on the intensity and the contour of the
target. The number of the segmentation labels is the same as
the one of the original images and when the original images
encounter the data augmentation, the segmentation labels also
go thought the data augmentation in the same way.

The size of the input SAR image is 80x80, the stride size
of each convolutional layer is set as 1x1, the batch size is 128, and the learning
rate is 0.0002, the epoch is set as 50. Some other hyper parameter are contained in
Fig.1. And the platform is tensorflow with with an Intel sandy bridge 4 core CPU at 3.5 GHz and a Nvidia 1080ti GPU.

\subsection{Segmentation Performance}
To verify the effectiveness of the proposed deep deformable residual learning network in SAR image segmentation visually, there are some segmentation results of different targets present in Fig.4. There are the original images, ground truth of segmentation, the corresponding masked original images, the segmentation results and the masked segmentation results from left to right in Fig.4.
It is clear that the segmentation results of the proposed network are almost same as the ground truth. The segmentation results greatly preserve the geometric features of the different targets compared to the ground truth.

To evaluate the segmentation results qualitatively, we employe the
pixel accuracy of the segmentation results. The pixel accuracy is calculated as follows
\begin{align}
\label{rangereso1}
{P_{pa}} = sum\left( {{P_p}} \right)/\left( {{P_a}} \right)
\end{align}
where ${P_{pa}}$ is the pixel accuracy, ${P_p}$ is the correct predicted pixel and ${P_a}$ is the total pixels in one SAR image. And it means that higher the pixel accuracy is, better the performance is.

\begin{table}[!htb]
\small
\centering
\begin{spacing}{1.4}
\caption{Pixel accuracies for the targets and the backgrounds(pixel accuracy 99.25\%).}
\begin{tabular}{p{2.3cm}<{\centering}p{2.3cm}<{\centering}p{2.3cm}<{\centering}}
\hline \hline
 Pixel accuracy & Target & Background  \\
\hline
Target & 99.52 & 0.48  \\
\hline
Background & 0.65 & 99.35 \\
\hline \hline
\end{tabular}
\end{spacing}
\label{para}
\end{table}

The pixel accuracy of the proposed multi-tasks deep learning
framework is present in the form of the confusion matrix
in Table II. From Table II, the overall accuracy
of the segmentation is higher than 99.00\%. It is clear that the proposed network
has the capability of segmenting the targets from
the backgrounds precisely and effectively.

From the evaluations of the performance of the target
segmentation, it can be proved that, through
deep deformable residual learning network, the target segmentation is obtained accurately
and effectively.

\section{CONCLUSION}
In general, the geometric features of target could provide unique information for SAR application.
In this paper, we proposed a new deep deformable residual learning network for SAR image segmentation.
By employing the deformable convolutional layers, the network obtains the capability of extracting and preserving the precise contour of the targets in the SAR images, and by employing and adopting the residual learning kernel though in the design of the network structure, the network has the capability of preserving the crucial information for the targets.
Extensive experiments are carried out on the
MSTAR data set, and the results show clearly that the proposed
deep deformable residual learning network could achieve superior segmentation performance and preserve the precise contour of the targets in SAR images.

\bibliographystyle{IEEEtran}
\bibliography{ref,ref_self}

\end{document}